\documentclass{svproc}
\usepackage{url}
\usepackage{graphicx}

\begin{document}
\mainmatter              % start of a contribution
\title{Singular Spectrum Analysis of Time-series Data from Time-dependent density-functional theory in Real-time}
\titlerunning{SSA for TDDFT time-series}  % abbreviated title (for running head)
%\author{Naoki Tani\inst{1} \and Satoru S. Kano\inst{1} \and Yasunari Zempo\inst{1} }
\author{Naoki Tani \and Satoru~S.~Kano \and Yasunari Zempo }
\authorrunning{Naoki Tani et al.} % abbreviated author list (for running head)
\tocauthor{Naoki Tani, Satoru~S.~Kano and Yasunari Zempo}
\institute{Hosei University, Computer and Information Sciences, Tokyo 182-8584, Japan,\\
\email{naoki.tani.7x@stu.hosei.ac.jp},\\
\texttt{https://cis.hosei.ac.jp/}}

\maketitle              % typeset the title of the contribution

\begin{abstract}
This paper introduces a spectral analysis of time-series data derived from real-time time-dependent density-functional theory (TDDFT) using Singular Spectrum Analysis (SSA). TDDFT is a robust method for obtaining molecular excited states and optical spectra by tracking the time evolution of dynamical dipole moments. However, the spectral resolution can be compromised when Fourier transform’s total time duration is insufficient. SSA enabled the extraction of specific oscillation components from the time-series data, facilitating the generation of higher-precision spectra. Even with relatively short time-series datasets, the predictive extension of SSA yielded high-resolution spectra, demonstrating substantial agreement with results obtained through conventional methods. The efficacy of this approach was validated for several small molecules, including ethylene, benzene, and others. SSA’s ability to conduct detailed spectral analysis in specific energy regions enhance spectral resolution and facilitates the clarification of oscillation components within these regions. Real-time TDDFT combined with SSA provides a new analytical method for analyzing the optical properties of molecules, significantly improving the accuracy of the analysis of emission and absorption spectra analysis. This method is expected to have various applications.

\end{abstract}

\section{Introduction} \label{Introduction}
Time-dependent density-functional theory (TDDFT) \cite{RG} is a powerful method for calculating the electronic structure and excited states of materials, and plays an important role in materials development and optical property analysis. In particular, real-time, real-space based TDDFT is widely used, due to its simplicity and intuitive operability \cite{YB}.

We employed real-time TDDFT to analyze the optical properties of organic light-emitting diodes and other devices \cite{YZNA}. This approach is easy to parallelize, enabling stable and efficient calculations. However, the accuracy and resolution of the calculated results are influenced by the length of the time evolution data. For example, when applying Fourier Transform (FT) to time-dependent dynamic dipole moment data to obtain absorption and emission spectra, insufficient total time can lead to reduced spectral resolution, making features like band-edge peaks indistinct. Since TDDFT’s are based on first-principles calculations, they demand significant computational resources. Inadequate computing resources or data may result in broadened spectra, complicating the accurate evaluation of optical properties. To solve this problem, we developed a new method for extracting useful information from short time-step data \cite{Togoshi2014,Togoshi2015,Togoshi2016}

This study proposes the application of Singular Spectrum Analysis (SSA) to real-time TDDFT results to extract essential oscillational components from short time-series data. By using SSA, fundamental oscillational components that contribute to the spectrum can be separated, allowing for a clearer analysis of band-edge peaks. Furthermore, by adding the predicted data to the insufficient time-series data based on the separated oscillation components and extending the effective total time, it is possible to achieve higher resolution spectral analysis.

%This paper aims to propose a novel spectral analysis method using SSA and demonstrate its effectiveness. Section 2 outlines the theoretical framework and methodology of real-time TDDFT and SSA. In section 3, we show the results of spectral analysis with SSA using several simple molecular systems. Finally, section 4 presents the conclusions of this study and discusses future prospects.

%
\section{Methods} \label{sec_TDDFT} %sec 2
\subsection{Time-dependent density-functional theory}  % subsec 2.1

In this section, we outline the real-time TDDFT calculation procedure. The total energy of the ground state, based on density-functional theory (DFT) with the local density approximation, is derived from the Kohn--Sham  (KS) equation. While DFT is inadequate for accurately describing optical responses and absorption spectra involving electronically excited states, this limitation was addressed by Runge and Gross through the introduction of time-dependent evolution of the DFT equation \cite{RG}. The equations of motion of TDDFT coupled with pseudopotentials can be written as follows:

\begin{eqnarray}
\begin{array}{c}
	i\frac{\partial }{{\partial t}}\psi _j ({\bf r},t) {\ } = H \psi _j ({\bf r},t) \\[2mm]
	H =
	- \frac{1}{2}\nabla ^2   
	+ V^{ps}_{ion} ({\bf r})
	+ V_{H} ({\bf r},t)
	+ V_{XC} [\rho ({\bf r},t)]
	+ V_{ext} ({\bf r},t) ,
\end{array}
\label{eq:tddft}
\end{eqnarray}
where $\psi_j$ represents the $j$ th wave function. Hamiltonian $H$ is the KS Hamiltonian $H_{KS}$ to which a perturbation $V_{ext}({\bf r},t)$ is added, i.e.,  $H=H_{KS}+V_{ext}({\bf r},t)$. $V_{ion}^{ps}$ represents the ionic pseudopotential, $V_{H}$ is the Hartree potential, and $V_{XC}$ is the exchange-correlation potential. The Hartree potential and the exchange-correlation potential are expressed by the electronic charge density $\rho({\bf r},t) = \sum_j |\psi_j({\bf r}, t)|^2$. The summation is performed over all occupied states $j$, and the Hartree potential is determined by $\nabla^2 V_H = -4\pi\rho$. For simplicity, all calculations in this study are performed using the atomic unit system.

Prior to the calculation of the optical response, we first determine the stationary state using conventional DFT to optimize the electronic structure. Next, we added the external potential $V_{ext}$ as a perturbation to the system and tracked the linear response of the system in real-time. In our calculations, we employ the real-time, real-space approach to solve Eq.(\ref{eq:tddft}) using the finite difference method. This approach is advantageous due to its suitability for parallel computation. In this study, a uniform grid is used for simplicity.

The external electric field $V_{ext}(t)$ is applied as a perturbation in the form $V_{ext}(t)=E(t)\xi$ in the $\xi$-direction at $t=0$. When a very weak electric field $E(t)$ is applied, a dipole moment $\mu_{\xi}(t)$ is generated as the system’s response. The polarizability, $\alpha_{\xi}(\omega)$, which  characterizes the linear response, is expressed as follows. Here, $\xi$ represents the $x,y,z$ directions.
\begin{equation}
	\int dt {\ } e^{i\omega t} {\ }\mu_{\xi}(t) = 
	\alpha (\omega ) \int dt {\ } e^{i\omega t} {\ }E(t).
	\label{eq:polarizability}
\end{equation}

When an external stimulus in the form of a delta function, $V_{ext}(t) = -k \xi \delta(t)$, is applied, the polarizability $\alpha_{\xi}(\omega)$ can be directly obtained from FT of the dipole moment $\mu_{\xi}(t)$
\begin{equation}
	\alpha_{\xi} = \frac{1}{k} \int dt {\ } e^{i\omega t} {\ }\mu_{\xi}(t) ,
	\label{eq:alpha_FT}
\end{equation}
where $k$ represents the strength of the external perturbation in the $\xi$ direction. Furthermore, the imaginary part of the polarizability is used to compute the total oscillator strength $S(\omega)$
\begin{equation}
	S(\omega ) = \frac{{2\omega }}{\pi }{\mathop{\rm Im}  \nolimits}
	{\ } \alpha (\omega ) ,
	\label{eq:osc_strength}
\end{equation}
where $\alpha = (\alpha_x + \alpha_y + \alpha_z)/3$ is the average polarizability. From Eq.(\ref{eq:tddft}), the time-dependent wave function $\psi_j({\bf r}, t)$ is expressed as follows:
\begin{equation}
	\psi_j ({\bf r},t) = e^{-iHt}\psi_j ({\bf r},0) .
	\label{eq:psi_evo}
\end{equation}

The initial wave function ${\tilde \psi}_{j}$ for this perturbation at $t=0$ is obtained from the following equation:
\begin{equation}
	{\tilde \psi} {\big|}_{\small t = 0}
	= \exp{ {\bigl [ } -i \int_{-0}^{+0} dt {\ }( H_{KS}-k\xi\delta(t) ) {\bigr ]} }
	{\ }\psi_j ({\bf r},0) .
	\label{eq:init_phi_integral}
\end{equation}
It becomes a very simple expression:
\begin{equation}
	{\tilde \psi} {\big|}_{\small t = 0}
	= e^{ik\xi} \psi_j ({\bf r},0) .
	\label{eq:init_phi} 	 
\end{equation}

The time evolution of the dipole moment, $\mu_{\xi}(t)$, is numerically computed up to a total time $T$ using discrete time steps of $\Delta t$, where $T=N\Delta t$. The spectral resolution of $S(\omega)$, obtained from Eq.(\ref{eq:osc_strength}), is approximately $O(1/T)$. Therefore, insufficient data points reduce spectral resolution, requiring a sufficiently large $T$ for a clear spectrum. To address this limitation, we propose applying SSA to the time-series $\mu_{\xi}(t)$ to enable clear spectral analysis even with limited data.

\subsection{Singular spectrum analysis} \label{sec_SSA} % subsec 2.2

The procedure for applying SSA to the time-series data $\mu_{\xi}(t)$ is described below. SSA is a method for decomposing time-series data and extracting components based on eigenvalues. 
A trajectory matrix is created from the time-series data, followed by Singular Value Decomposition (SVD).
The resulting components were then reconstructed to identify the fundamental oscillation components in the data, which were analyzed using FT \cite{Golyandina,Brunton}.

From the time-series data $F = (f_1, f_2, \dots , f_N)$, a trajectory matrix $Y$ of size $\tau \times n$ is created using window width $\tau (1<\tau <N/2)$, where $n=N-\tau+1$. This matrix is skew-symmetric and captures the correlation in the original time-series data:

\begin{eqnarray}
Y=
\left(
\begin{array}{cccc}
	f_1       & f_{2}     & \cdots & f_{n} \\
	f_2       & f_{3}     & \cdots & f_{n+1} \\
	\vdots    & \vdots    & {\ }   & \vdots  \\
	f_{\tau}  & f_{\tau+1}& \cdots & f_{N} \\	
\end{array}
\right).
% =({\bf y}^{(1)},{\bf y}^{(2)},\dots,{\bf y}^{(n)}),
\label{eq:ts_org}
\end{eqnarray}

The column vectors of $Y$ are expressed as ${\bf y}^{(1)},{\bf y}^{(2)},\dots,{\bf y}^{(n)}$. SVD is applied to this matrix, decomposing it into $Y = U \Sigma V^T$, where $U$ is a $\tau \times \tau$ matrix with column vectors $({\bf u}_1,\dots,{\bf u}_{\tau} )^T$. $\Sigma$ is a diagonal matrix of size $\tau \times n$, and each diagonal component is given by $(\sigma_1, \dots, \sigma_{\tau})$. $V^{T}$ is also a $n \times n$ orthogonal matrix with row vectors $({\bf v}_1,\dots,{\bf v}_{\tau} )^T$, respectively. To remove high-frequency noise, low-rank approximations are used in SVD. Based on the singular values, $Y$ is decomposed into its cor-responding components $Y_i$:
\begin{equation}
Y \approx U_r \Sigma_r V_r^T ={U'}_r V_r^T = \sum_{i=1}^{r} Y_i,
\label{eq:svd}
\end{equation}
where $r$ is the adopted rank. For simplicity, the product of $U_r$ and $\Sigma_r$ is defined as ${U'}_r$. The $i$ th component $Y_i$ can be expressed as
\begin{equation}
	Y_i=
	\left( {\bf y}_i^{(1)}, {\bf y}_i^{(2)}, \dots, {\bf y}_i^{(n)}
	\right) .
	\label{eq:ts_dc_each}
\end{equation}

In general, this matrix $Y_i$ is not skew-symmetric like the original matrix $Y$. Using the average as shown in Eq.(\ref{eq:fave}), we reconstruct the following time-series data ${\tilde F}_i=({\tilde f}_{i,1},{\tilde f}_{i,2},\dots,{\tilde f}_{i,N})$, which corresponds to each singular value $\sigma_i$:
\begin{eqnarray}
{\tilde f}_{i,s} =
\left \{
\begin{array}{ccc}
	\frac{1}{s} \sum\limits_{j=1}^{s} y_{i{\tiny(j,s-j+1)}}              &{\ }& (1\le s \le \tau)   \\
	\frac{1}{\tau} \sum\limits_{j=1}^{\tau} y_{i{\tiny (j,s-j+1)}}       &{\ }& (\tau \le s \le n )  \\
	\frac{1}{N-s+1} \sum\limits_{j=1}^{N-s+1} y_{i{\tiny (j+s-n,n-j+1)}} &{\ }& (n \le s \le N) .  
	\\
\end{array}
\right.
\label{eq:fave}
\end{eqnarray}
Consequently, we can obtain a matrix ${\tilde Y}_i$
\begin{eqnarray}
{\tilde Y}_i
= \left(
\begin{array}{cccc}
	\tilde{f}_{i,1}       & \tilde{f}_{i,2}     & \cdots & \tilde{f}_{i,n} \\
	\tilde{f}_{i,2}       & \tilde{f}_{i,3}     & \cdots & \tilde{f}_{i,n+1} \\
	\vdots            & \vdots            & {\ }   & \vdots  \\
	\tilde{f}_{i,\tau}  & \tilde{f}_{i,\tau+1}& \cdots & \tilde{f}_{i,N} \\	
\end{array}
\right) .
\label{eq:reconstructedY}
\end{eqnarray}

The reconstructed matrix ${\tilde Y}_i$ retains symmetry similar to that of the original trajectory matrix $Y$. Now we also place the column vector ${\tilde Y}_i$ as ${\tilde {\bf y}}_i^{(1)},
{\tilde {\bf y}}_i^{(2)},\dots, {\tilde {\bf y}}_i^{(n)}$. The original time-series can be approximated as $F\approx \Sigma_i {\tilde F}_{i}$.

In some cases, multiple $\tilde{F}_{i}$ components may exhibit oscillations of similar frequency and amplitude. These components are grouped into the same cluster and treated as a single oscillation. To classify these components into clusters, the similarity of their oscillation behavior was evaluated using the W-correlation matrix \cite{SSAcode}
\begin{equation}
W_{i,j} = \frac{(\tilde{F}_i,\tilde{F}_j)}{\|{\tilde{F}}_i\|{\ }\|{\tilde{F}}_j\|}.
\label{eq:wij}
\end{equation}

\subsection{Forecast of the time-series} \label{sec_forecast} %subsection 2.3

This section explains the forecasting of the time-series data $\tilde{F}_i$ using SSA \cite{Danilov}. As mentioned above, since the time-series data $\tilde{F}_i$, extracted as specific oscillation components, is significantly simpler than the original $F$, it allows for stable forecasting. Forecasting $\tilde{F}_{i,N+1}$ using $\tilde{F}_i = (\tilde{f}_{i,1},\tilde{f}_{i,2},\dots,\tilde{f}_{i,N})$ involves adding a new column vector of $\tilde{Y}_i$ as shown Eq.(\ref{eq:reconstructedY}).
\begin{eqnarray}
{\bf \tilde{y}}_i^{(n+1)}=
\left(
\begin{array}{c}
	{\tilde f}_{i,n+1}      \\
	\vdots   \\
	{\tilde f}_{i,N}  \\
	{\tilde f}_{i,N+1}  \\	
\end{array}
\right).
\label{eq:forecast_vec}
\end{eqnarray}

Since only ${\tilde f}_{i,N+1}$ is unknown, we use the first to the $\tau-1$ of ${\tilde Y}_i$, and obtain the following linear combination of the components ${\tilde u}_{i,jk}$ with coefficients $h_k$ as follows:
\begin{eqnarray}
	\displaystyle
	{\tilde f}_{i,n+1} =&
	\frac{1}{n+1} \sum\limits_{s=1}^{n+1} {\tilde f}_{i,s}&
	+	\sum\limits_{k=1}^{r} h_k {\tilde u}_{1,k},   \nonumber\\ %[1mm]
	\label{eq:ls_forecast}
	&\vdots&   \\[1mm]
	\displaystyle
	{\tilde f}_{i,n+\tau-1} =&
	\frac{1}{n+\tau-1} \sum\limits_{s=\tau-1}^{n+\tau+1} {\tilde f}_{i,s} &
	+	\sum\limits_{k=1}^{r} h_k {\tilde u}_{\tau-1,k},  \nonumber %\\[1mm] \nonumber
	%	\end{array} 
\end{eqnarray}
where ${\tilde u}_{j,k}~(j=1,\dots,\tau-1)$ are the components of ${U'}_{r}^{(i)}$, obtained from the decomposition  as per Eq.(\ref{eq:svd}). The first term in Eq.(\ref{eq:ls_forecast}) is the average of each row $f_{i,a}~(a=1,\dots,\tau-1)$ of ${\tilde Y}_{i}$, and the second term is the deviation of the component. The new point, ${\tilde f}_{i,N+1}$, is obtained by solving Eq.(\ref{eq:ls_forecast}) for $h_k$ using the least-squares method. The forecast is made by
\begin{equation}
{\tilde f}_{i,N+1} = \frac{1}{n} \sum\limits_{s=\tau}^{N} {\tilde f}_{i,s}
+ \frac{n+1}{n} \sum\limits_{k=1}^{r} h_k {\tilde u}_{\tau,k}.
\label{eq:forecast}
\end{equation}

By incorporating this value into ${\tilde Y}_{i}$ as a new component to ${\bf \tilde{y}}_i^{(n+1)}$, an updated ${\tilde Y}_i$ is obtained, expanded by one column. SVD is then reapplied to the updated ${\tilde Y}_i$, allowing the operations described in Eq.(\ref{eq:ls_forecast}) and (\ref{eq:forecast}) to be repeated iteratively, enabling forecasts to be extended to the desired number of points.

\subsection{Estimated computational cost} \label{sec_comp_cost} %subsection 2.4
The majority of the computational cost for the proposed method arises from the time-series data generated by real-time TDDFT calculations.
In real-space real-time TDDFT, the orbital orthogonalization --- which scales with the number of orbitals, proportional to the number of $M$ --- is the rate-limiting step.
As a result, the computational complexity scales as $O(M^2N)$, where $N$ is the number of time steps.

In contrast, FT and SSA using SVD depend only on the number of time-series data points $N$ and are independent of the system size.
Consequently, these operations are computationally orders of magnitude lighter than TDDFT.
Specifically, SSA requires the SVD of a $\tau \times n$ matrix $(\tau<n)$ truncated to rank $r$,
resulting in a computational complexity of $O(\tau n r)$.
When performing the point-by-point prediction described in Section \ref{sec_forecast} $L$ times,
the total computational cost can be estimated as $O(\tau r L N + \tau r L^2)$.
Therefore, even on a standard desktop computer, the spectral analysis and prediction steps can be efficiently executed.

\section{Results and discussion}
\subsection{Ethylene} \label{sub_ethylene}
\begin{figure}[h!]
\begin{minipage}{\linewidth}%\hspace{1pc}
	\includegraphics[width=\linewidth]{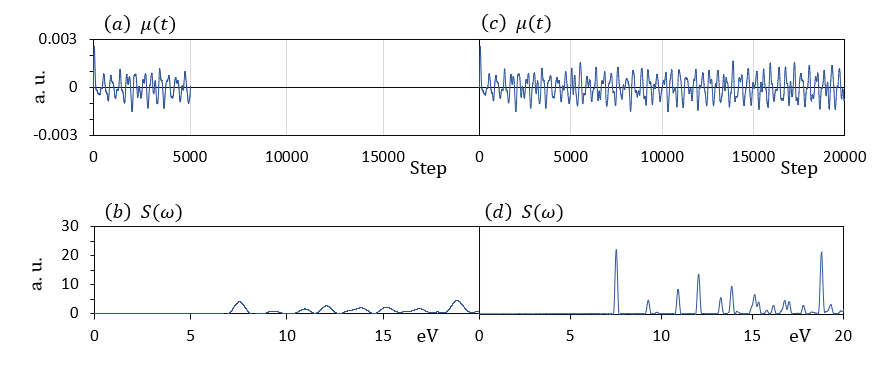} %\hspace{5pc}%
	\end{minipage}\hspace{1pc}
%	\begin{minipage}{16pc}
	\begin{minipage}{\linewidth}
		\caption{\label{fig:dp_S} Dipole moment $\mu(t)$ and oscillator strength $S(\omega)$ for steps ({\it a})({\it b}) $N=5000$ and ({\it c})({\it d}) $N=20000$, respectively.}
        \end{minipage}
\end{figure}
To evaluate the effectiveness of SSA, we applied it to the dipole moments obtained from TDDFT calculations for ethylene, a molecule with well-characterized electronic states. 
The initial external stimulus was applied along the C-C bond direction, resulting in a slowly varying dipole moment $\mu(t)$. Spectral analysis of $\mu(t)$ was performed using FT as described in Eq.(\ref{eq:alpha_FT}), focusing on the band-edges associated with optical absorption. 
Figure \ref{fig:dp_S} shows the dipole moments and the resulting spectrum $S(\omega)$ for two cases: (1) sufficient data ($N\simeq20000$) and (2) insufficient data ($N\simeq5000$), with a time step$\Delta t=0.002$~[1/eV]. When sufficient time evolution is available, the band-edge spectra are clear, with distinct peaks around $7.5$ eV. 
In contrast, with insufficient data points ($N\simeq5000$), the spectrum becomes weak and broad, making it challenging to discern whether it is a single peak or composed of multiple oscillations. 
Identifying peaks associated with the band-edges also becomes difficult. We then analyzed the spectra using the shorter time-evolved time-series $F=(\mu_1, \mu_2, \dots, \mu_{5000})$ up to $N=5000$ steps. 
It is necessary to select the appropriate oscillation components related to the band-edge among several obtained by decomposing the time-series $F$. 
To identify the major oscillation components in $F$, a trajectory matrix $Y$ is constructed with the window width $\tau=1000$ according to Eq.(\ref{eq:ts_org}). 
The results are obtained in the order of the largest singular values. Using $Y_1,Y_2,\dots,Y_r$ according to Eq.(\ref{eq:ts_dc_each}), and by averaging according to Eq.(\ref{eq:fave}), each time-series data ${\tilde F}_1,{\tilde F}_2,\dots$ corresponding to each singular value is obtained.
\begin{figure}[h]
\begin{center}
\begin{minipage}{0.9\linewidth}%\hspace{1pc}
\includegraphics[width=\linewidth]{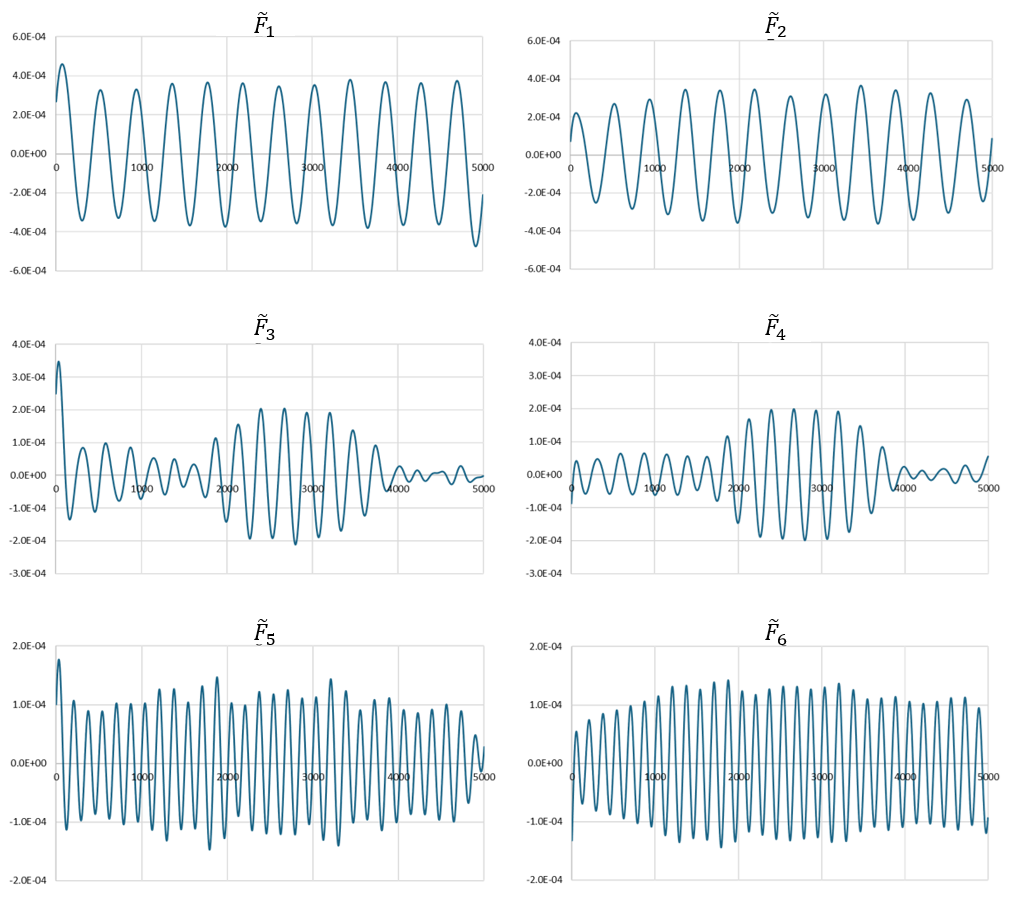} %\hspace{5pc}%
\end{minipage}\hspace{1pc}
\end{center}
%	\begin{minipage}{16pc}
\begin{minipage}{\linewidth}
\caption{\label{fig:decomp} Oscillations of the dipole moment $\mu(t)$ for ethylene were decomposed and reconstructed into individual signal components using SSA with a bandwidth $\tau=1000$.}
\end{minipage}
\end{figure}

Figure \ref{fig:decomp} shows the first six oscillations. Pairs such as (${\tilde F}_1$,${\tilde F}_2$), (${\tilde F}_3$,${\tilde F}_4$), and (${\tilde F}_5$,${\tilde F}_6$) exhibit similar frequencies and amplitudes, indicating they are related oscillations. Each pair was clustered and denoted as $({\tilde F}_i,{\tilde F}_j)$. For these reconstructed oscillations, we counted the wavenumbers, excluding the first and the last approximately $100$ steps to avoid boundary effects. The spectra derived from this wavenumber analysis showed that the main oscillations correspond to peaks around $7.5$~eV for $({\tilde F}_1,{\tilde F}_2)$, $11.8$~eV for  $({\tilde F}_3,{\tilde F}_4)$ and $18.4$~eV for  $({\tilde F}_5,{\tilde F}_6)$. Comparing these results with Fig.\ref{fig:dp_S}({\it b}) confirms that these peaks are included.

The peaks near the corresponding energies are notable. In particular, $({\tilde F}_1,{\tilde F}_2)$ represent the lowest energy peaks, which are associated with the band-edge. These oscillations exhibit minimal beating and large amplitudes, indicating that they are relatively simple signals. In contrast, $({\tilde F}_3,{\tilde F}_4)$ and $({\tilde F}_5,{\tilde F}_6)$ show noticeable beats and are mixed with adjacent oscillations. This indicates that SSA does not always achieve a perfect decomposition into single-frequency components.
%\vspace{-3mm}

\begin{figure}[h]
\begin{minipage}{0.5\linewidth}%\hspace{1pc}
\vspace{-4mm}
%\hspace{10mm}\includegraphics[width=0.5\linewidth]{Figs/wij.png} %\hspace{5pc}%
\hspace{10mm}\includegraphics[width=0.5\linewidth]{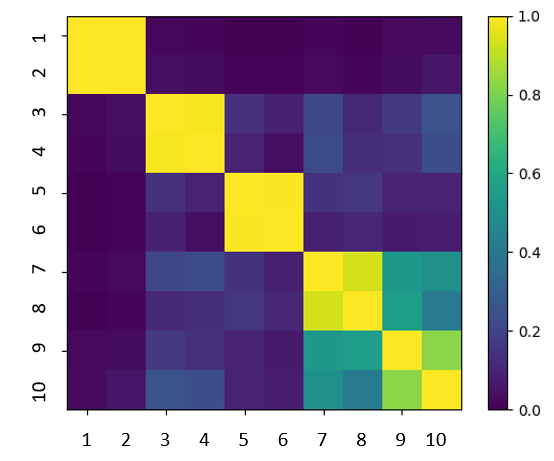} %\hspace{5pc}%
\vspace{-5mm}
\end{minipage}%\hspace{1pc}
%	\begin{minipage}{16pc}
\begin{minipage}{0.45\linewidth}
	\vspace{-4mm}
	\caption{\label{fig:wij} Correlation matrix on the decomposed and reconstructed time-series data $\{{\tilde F}_1,\dots,{\tilde F}_{10} \}$ of ethylene dipole moment.}
	\vspace{-5mm}
\end{minipage}
\end{figure}

Figure \ref{fig:wij} shows the correlation matrix $W_{ij}$ for ${\tilde F}_1 - {\tilde F}_{10}$, calculated using Eq.(\ref{eq:wij}). 
The diagonal components are, as expected, equal to $1$, as they represent self-correlation. 
If different time-series are completely separated, their off-diagonal correlations vanish. 
However, Fig.(\ref{fig:wij}) shows high correlations within the pairs $({\tilde F}_1,{\tilde F}_2)$,$({\tilde F}_3,{\tilde F}_4)$, and $({\tilde F}_5,{\tilde F}_6)$, confirming their relationship. 
The effectiveness of this decomposition depends on the window width $\tau$, which must be optimized for proper separation. In SSA, the outputs are ordered from the highest singular value to the lowest, which means that the main oscillations may appear in the high-energy region, depending on the direction of the dipole moment oscillation. To analyze the band-edges effectively, it is useful to rearrange the decomposed and reconstructed time-series data in ascending order of wavenumber, renaming them as ${\tilde F}_1,{\tilde F}_2,\dots$, for clarity.

\begin{figure}[h!]
\begin{center}
	\begin{minipage}{0.9\linewidth}%\hspace{TDDFT1pc}
		\hspace{15mm}\includegraphics[width=0.7\linewidth]{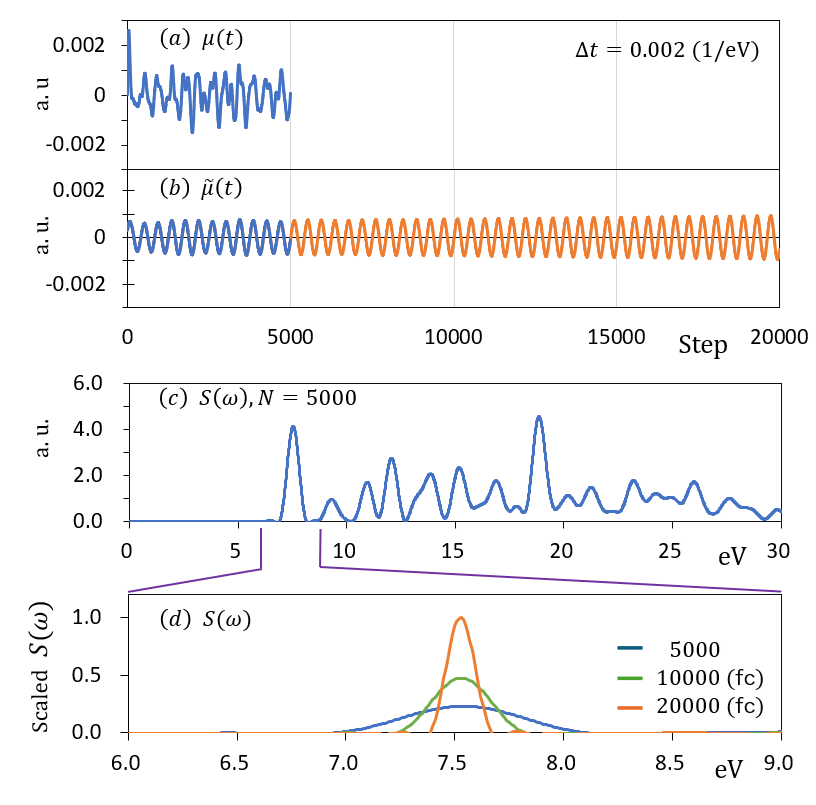}	
%      \hspace{15mm}\includegraphics[width=0.7\linewidth]{Figs/forecast.png} %\hspace{5pc}%
	\end{minipage}\hspace{1pc}
\end{center}
%	\begin{minipage}{16pc}
	\begin{minipage}{\linewidth}
		\caption{\label{fig:forecast} Dynamic dipole moment and spectrum of ethylene. ({\it a}) the base $\mu(t)$ 
			and ({\it c}) its spectrum, ({\it b}) the extraction of the band-edge component by SSA and its spectral prediction $\mu(t)$, and ({\it d}) its spectral improvement,
			normalized by the maximum of the peak obtained from $N=20000$.
            Dynamic dipole moment and spectrum of ethylene. ({\it a}) the base $\mu(t)$ and ({\it c}) its spectrum, ({\it b}) the extraction of the band-edge component by SSA and its spectral prediction$\mu(t)$, and ({\it d}) its spectral improvement, normalized by the maximum of the peak obtained from $N=20000$.}
	\end{minipage}
	\vspace{-5mm}
\end{figure}

Figure \ref{fig:forecast} demonstrates the analysis and forecasting of oscillations using the method described in Section \ref{sec_forecast}. 
In Fig.\ref{fig:forecast}({\it a}), the time-series data $F=(\mu_1,\mu_2,\dots,\mu_{5000})$ is shown for the dynamic dipole moment $\mu(t)$ up to $N=5000$ steps. 
Fig.\ref{fig:forecast}({\it b}) displays the time-series data ${\tilde F}={\tilde F}_1+{\tilde F}_2$, and the forecasted oscillation data using the prediction method from Section \ref{sec_forecast}. In the figure, ${\tilde \mu}(t)$ represents ${\tilde F}=({\tilde \mu}_1,{\tilde \mu}_2,\dots,{\tilde \mu}_{5000})$. For this ${\tilde F}$(blue), the prediction is repeated up to $N=20000$ steps, and the result is shown in orange. The reconstructed oscillation remains almost simple and stable over a long period of time. Figure.\ref{fig:forecast}({\it c}) shows the spectrum obtained by applying FT directly to the original time-series $F$ from Fig.\ref{fig:forecast}({\it a}). Since time-series up to $N=5000$ steps are used, the total time $T$ is not sufficient, and the spectrum is very broad with low resolution. Focusing on the low-energy peak, the spectrum in the energy region from $6$~eV to $9$~eV is shown in Fig.\ref{fig:forecast}({\it d}), with each peak normalized to the magnitude obtained from the results at $N=2000$ steps (orange). The spectra compare the FT results for ${\tilde F}$ using data from $N=5000$ steps (blue), $N=10000$ steps (green), and $N=20000$ steps (orange). It is evident that the signal becomes stronger and sharper as the number of steps increases.

\subsection{Small molecules (benzene / naphthalene / anthracene / tetracene)} \label{sec_small}
\begin{figure}[h!]
\begin{center}
	\begin{minipage}{1\linewidth}%\hspace{1pc}
		\includegraphics[width=\linewidth]{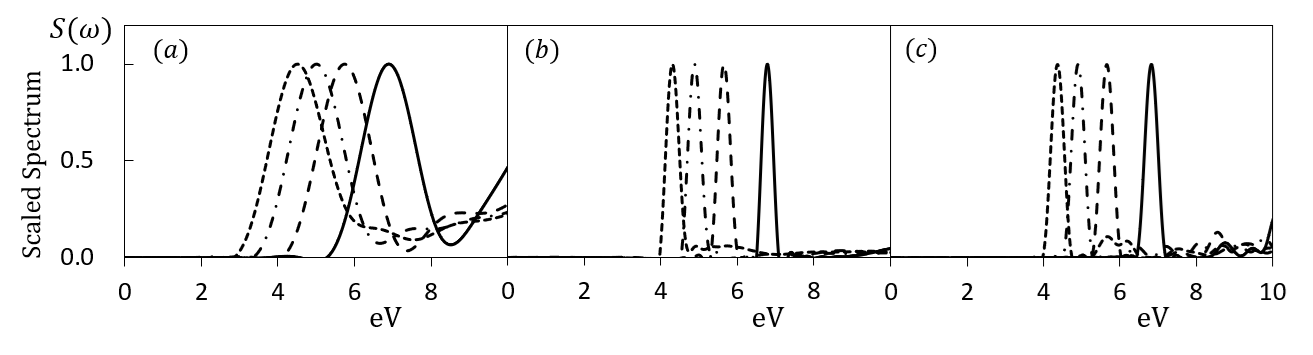} %\hspace{5pc}%		
	\end{minipage}\hspace{1pc}
\end{center}
%	\begin{minipage}{16pc}
	\begin{minipage}{\linewidth}
		\caption{\label{fig:smallmol}
            Comparison of the calculated band-edge spectra for benzene (solid line), naphthalene (dashed line), anthracene (dash dot line), and tetracene (dotted line). ({\it a}) Spectrum obtained from the original $\mu(t)$ calculated using TDDFT up to $N=2000$ steps. ({\it b}) Spectrum derived from the extracted oscillation ${\tilde \mu}(t)$, associated with the band-edge, extended to $N=8000$ steps using SSA and forecasting. ({\it c}) Spectrum obtained from the time evolution of up to $N=8000$ steps simply using TDDFT with $\Delta t=0.002$ [1/eV].
            }
	\end{minipage}
\end{figure}
In this section, we confirm the effect of the SSA forecast by comparing our results with those of TDDFT up to the same number of steps, using small molecules.
For molecules with simple electronic structures, such as benzene, naphthalene, anthracene, and tetracene, we examined the effect of forecasting on the spectra near the band-edge, as in Section \ref{sub_ethylene} It is well-known that the energy of the peak positions decreases as the molecular size increases.

Figure \ref{fig:smallmol} compares spectra derived from the dynamic dipole moments calculated by TDDFT with those extended and forecasted using SSA. Fig.\ref{fig:smallmol}({\it a}) shows the spectra near the band-edge, based on the dynamic dipole moment $\mu(t)$ obtained by TDDFT up to the step number ($N = 2000$). The results reveal a shift of the band-edge to lower energy as the number of benzene rings and atomic size increase. 
However, the spectrum in this energy region appears ambiguous, influenced by other nearby peaks.

To clarify the band-edge spectrum, we applied SSA to extract ${\tilde \mu}(t)$ from each time-series data at $N=2000$ steps, isolating the oscillations associated with the band-edge. 
The extracted fundamental oscillation was then forecasted and extended to $N=8000$ steps, as shown in Fig.\ref{fig:smallmol}({\it b}), with a window width of $\tau=500$. 
Since SSA effectively isolates the oscillation components near the band-edge, the influence of other peaks is relatively reduced. 
On the other hand, Fig.\ref{fig:smallmol}({\it c}) shows spectra obtained from the dipole moments calculated up to the step number ($N=8000$) simply by the TDDFT calculation.

Comparing the peak shapes of these spectra, it is found that they are nearly identical and equivalent.

\section{Conclusion} \label{conclusion}

Real-time TDDFT is a powerful method for obtaining the excited states and optical spectra of molecules. Spectra across all energy regions can be derived by applying the FT to the dynamic dipole moment. However, this approach assumes that the total time of the time evolution is sufficiently long. When the total time is insufficient, the spectral resolution is reduced, leading to broad spectral features. Consequently, the conventional FT method struggles with broad and ambiguous spectral shapes when the number of data points is limited. In addition, the dynamical dipole moment contains multiple frequency components, complicating the analysis. Recognizing that the dynamical dipole moment is time-series data, we applied SSA to focus on specific oscillation components within the dipole moment. By decomposing the dynamical dipole moment into groups of simpler oscillations, we were able to isolate and extract oscillations associated with specific spectral peaks. This approach allowed us to identify low-energy oscillations, especially those at the band-edges related to emission and absorption, and to analyze the spectra of individual oscillation components.

This decomposition provides relatively simple oscillation components within specific energy regions. Leveraging these components, we can forecast and extend the oscillations, effectively increasing the total time available for FT. As a result, the spectral shapes became very clear, yielding sharp and high-resolution spectra for specific energy regions. Importantly, the spectra obtained from the forecasted and extended time-series showed excellent agreement with those obtained from real-time TDDFT calculations using sufficiently long time-series data. These results were validated through analysis of ethylene and small molecules such as benzene, naphthalene, anthracene, and tetracene. This demonstrates the utility of SSA in enhancing the spectral analysis of real-time TDDFT calculations, particularly for detailed investigations of specific energy regions.

%\ack
\section*{Acknowledgment}
This work was partially supported by the Takahashi Industrial and Economic Research Foundation, and Sumitomo Chemical Co., Ltd.


\end{document}